\documentstyle[prl,aps,epsf,array]{revtex}
\begin{document}
\draft

\title{Universality in metallic nanocohesion: a quantum chaos approach}

\author{C.\ A.\ Stafford,$^{1,2,3}$ F.\ Kassubek,$^{1,2,3}$ 
J.\ B\"urki,$^{1,4,5}$ and Hermann Grabert$^3$}
\address{$\mbox{}^1$Physics Department, 
University of Arizona, 1118 E.\ 4th Street,
Tucson, AZ 85721}
\address{$\mbox{}^2$Institute for Theoretical Physics, University of
California, Santa Barbara, CA 93106-4030}
\address{$\mbox{}^3$Fakult\"at f\"ur Physik, Albert-Ludwigs-Universit\"at,
Hermann-Herder-Stra\ss e 3, D-79104 Freiburg, Germany}
\address{$\mbox{}^4$Institut de Physique Th\'eorique, Universit\'e de
Fribourg, CH-1700 Fribourg, Switzerland}
\address{$\mbox{}^5$Institut Romand de Recherche Num\'erique en
Physique des Mat\'eriaux, CH-1015 Lausanne, Switzerland}

%\date{}
\twocolumn[\hsize\textwidth\columnwidth\hsize\csname@twocolumnfalse\endcsname

\maketitle

\begin{abstract}

Convergent semiclassical trace formulae for the 
density of states and
cohesive force 
of a narrow constriction in
an electron gas, whose classical motion is either chaotic or
integrable, are derived.  It is shown that mode quantization
in a metallic point contact or nanowire leads to universal 
oscillations in its cohesive force:
the amplitude of the oscillations depends only on a dimensionless quantum
parameter describing the crossover from chaotic to integrable motion, and
is of order 1 nano-Newton, in agreement
with recent experiments.  
Interestingly, quantum tunneling is shown to be
described quantitatively in terms of the instability of the classical
periodic orbits.

\end{abstract}

%\pacs{PACS numbers: 
%73.40.Jn,  %Metal-to-metal contacts
%03.65.Sq,  %Semiclassical theories and applications
%05.45.Mt,  %Quantum chaos
%73.23.-b   %Mesoscopic systems
%}
\vskip2pc]

An intriguing question posed by Kac \cite{kac} is, ``Can one hear the shape of
a drum?'' That is, given the spectrum of the wave equation \cite{kac} or
Schr\"odinger's equation for free particles \cite{yanez} on a domain, can one
infer the domain's shape?  This question was answered in the negative
\cite{kac,yanez}; nevertheless there is an intimate relation
between the two.
In the context of metallic nanocohesion
\cite{nanoforce,sbb,comment.zwerger,blom,landman2,wnw,zwerger2,jerome}, 
a related question has recently emerged:
``Can one feel the shape of a metallic nanocontact?'' 
It was shown experimentally \cite{nanoforce} 
that the cohesive force of Au nanocontacts exhibits mesoscopic oscillations
on the nano-Newton scale, which are synchronized with steps of order 
$2e^2/h$ in the contact conductance.  In a previous article \cite{sbb},
it was argued that these mesoscopic force oscillations, like the corresponding
conductance steps \cite{beenakker},
can be understood by considering the nanocontact as
a waveguide for the conduction electrons (which are responsible for both 
conduction and cohesion in simple metals). Each quantized mode transmitted
through the contact contributes $2e^2/h$ to the conductance \cite{beenakker}
and a force of order $\varepsilon_F/\lambda_F$ to the cohesion, 
where $\lambda_F$ is the 
de Broglie wavelength at the Fermi energy $\varepsilon_F$.  It was shown
by comparing various geometries \cite{sbb}
that the force oscillations were determined by the area and symmetry 
of the narrowest cross-section of the contact, and depended only weakly
on other aspects of the geometry.  Subsequent studies 
confirmed this observation, both for generic geometries 
\cite{comment.zwerger,landman2,wnw,jerome}, whose classical dynamics is
{\em chaotic}, and for special geometries \cite{blom,zwerger2}, whose 
classical dynamics is {\em integrable}.  The insensitivity of the force
oscillations to the details of the geometry, along with 
the approximate independence of their r.m.s.\ size on the contact area, was
termed {\em universality} in Ref.\ \onlinecite{sbb}.
A fundamental explanation of the universality observed in both the 
model calculations 
\cite{sbb,comment.zwerger,blom,landman2,wnw,zwerger2,jerome}  
and the experiments \cite{nanoforce} has so far been lacking.

\begin{figure}
\epsfxsize=10cm
\epsffile{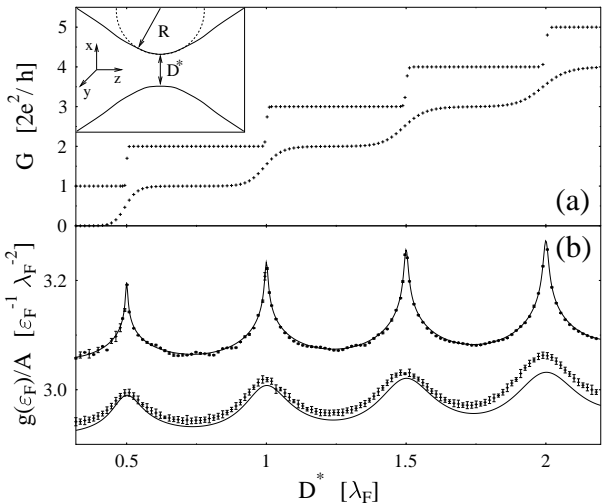}
\caption{
Inset: Schematic diagram of a metallic nanocontact.  (a) Conductance $G$ 
 and (b) DOS $g(\varepsilon_F)$ 
for 2D nanocontacts with $\alpha \approx 5$ 
versus the contact diameter $D^*$.   
$g$ is normalized
to the area $A$ of the region.
Solid curves: semiclassical result based on Eq.\ (\ref{dos.int.2d});
crosses with error bars: 
numerical results using the method of Ref.\ [10]. %\onlinecite{jerome}.
Lower curves in (a) and (b): $R\approx \lambda_F$; upper curves
(offset vertically): 
$R\approx 170\lambda_F$.
}
\label{fig.2}
\end{figure}

In this Letter, we derive semiclassical trace formulae 
for the force and charge oscillations of a metallic nanocontact, modeled as
a constriction in an electron gas with hard-wall boundary conditions
(see Fig.\ \ref{fig.2} inset), by adapting methods from quantum chaos
\cite{gutzwiller,brack,b&b.wigner,gut.symmetry,gut.interpolate}
to describe the quantum mechanics of such an open system.  It is found that
Gutzwiller-type trace formulae 
\cite{gutzwiller,brack,b&b.wigner,gut.symmetry,gut.interpolate}, 
which typically 
do not converge for closed systems, not only converge, but give quantitatively
accurate results for open quantum mechanical systems, which are typically 
more difficult to treat than closed systems by other methods.
Using these techniques, we demonstrate analytically that the force
oscillations $\delta F$ of a narrow constriction in a
three-dimensional (3D) 
electron gas (i) 
depend only on the diameter
$D^{\ast}$ and radius of curvature $R$ of the neck, 
(ii) have an r.m.s.\ value which is
independent of the conductance $G$ of the contact and depends
only on a scaling parameter $\alpha$ which describes the crossover from
chaotic to integrable motion, and (iii) are proportional to the charge
oscillations induced on the contact by the quantum confinement.
Furthermore, we show (iv) that quantum tunneling
through the constriction is determined by the instability of the classical
periodic orbits within the constriction, and that the force and charge
oscillations are suppressed only weakly (algebraically) by tunneling, unlike
conductance quantization, which is suppressed exponentially \cite{beenakker}.
Conclusion (ii) is specific to 3D
contacts, and breaks down for, e.g., two-dimensional (2D) nanowires, where
$\mbox{rms}\,\delta F \propto G^{-1/2}$.  
Conclusions (i), (ii), and (iv) 
are unchanged when electron-electron interactions are 
included within the Hartree approximation. 

The properties of simple metals are determined largely by the conduction 
electrons, the simplest model of which is a free-electron gas confined within 
the surface of the metal. 
Here we take the confinement potential
to be a hard wall; the effects of interest to us are virtually unchanged
when one considers a more realistic confinement potential \cite{landman2}.
The grand canonical potential $\Omega$ 
is the appropriate thermodynamic potential
describing the energetics of the electron gas in the nanocontact \cite{sbb},
and is
\begin{equation}
\Omega = -\frac{1}{\beta} \int dE\, g(E) \ln \left(1+e^{-\beta(E-\mu)}\right),
\label{ener.total}
\end{equation}
where $g(E)$ is the electronic density of states (DOS) and $\beta$ is the
inverse temperature \cite{domega}. 
The total number of electrons in the system is
\begin{equation}
N_- = \int dE\, f(E) g(E),
\label{charge.total}
\end{equation}
where $f(E)$ %=\{\exp[\beta(E-\mu)]+1\}^{-1}$ 
is the Fermi-Dirac distribution function.
The DOS of an open quantum system, such as that shown in Fig.\ \ref{fig.2}
(inset),
is given in terms of the electronic scattering matrix $S(E)$ by \cite{dashen}
$g(E) = (2\pi i)^{-1} \mbox{Tr} \{S^{\dagger}(E) \partial S /
\partial E - \mbox{H.c.}\}$,
where a factor of 2 for spin has been included.

The DOS 
can be decomposed \cite{brack,b&b.wigner} 
in terms of a smooth Weyl contribution $\bar{g}(E)$ 
and a fluctuating term $\delta g(E)$, 
\begin{equation}
g(E) = 
\frac{k_E^3{\cal V}}{2\pi^2 E} 
- \frac{k_E^2{\cal S}}{8\pi E}
+ \frac{k_E{\cal C}}{6\pi^2 E}
+ \delta g(E),
\label{dos.weyl}
\end{equation}
where $k_E=(2mE/\hbar^2)^{1/2}$,
${\cal V}$ is the volume of the electron gas, ${\cal S}$ is its surface 
area, and 
${\cal C} = \frac{1}{2}\int d\sigma\,\left(1/R_1+1/R_2\right)$ 
is the mean curvature of its surface, $R_{1,2}$ being the principal
radii of curvature.  The first three terms in Eq.\ (\ref{dos.weyl}) are 
macroscopic, while $\delta g$ determines the mesoscopic 
fluctuations of the equilibrium 
properties of the system.
Inserting Eq.\ (\ref{dos.weyl}) into Eqs.\ 
(\ref{ener.total}) and (\ref{charge.total}), and 
taking the limit of zero temperature, one finds
\begin{equation}
\frac{\Omega}{\varepsilon_F}  =  
-\frac{2k_F^3 {\cal V}}{15\pi^2} + \frac{k_F^2 {\cal S}}{16\pi} -
\frac{2k_F {\cal C}}{9\pi^2} + \frac{\delta \Omega}{\varepsilon_F},
\label{ener.weyl}
\end{equation}
\begin{equation}
N_-  =  \frac{k_F^3 {\cal V}}{3\pi^2} - \frac{k_F^2 {\cal S}}{8 \pi}
+ \frac{k_F {\cal C}}{3\pi^2} + \delta N_-,
\label{charge.weyl}
\end{equation}
where 
$k_F=2\pi/\lambda_F$ is the Fermi wavevector.
The corrections to Eqs.\ (\ref{ener.weyl}) and (\ref{charge.weyl})
at finite temperature
may be evaluated straightforwardly \cite{brack}, and are quite small
at room temperature, since $\varepsilon_F/k_B > 10^4K$.

The cohesive force of the nanocontact is given by the derivative of the
grand canonical potential with respect to the elongation,
$F=-\partial \Omega/\partial L$.
Under elongation, the contact narrows and its surface area ${\cal S}$ 
increases.  
The
increase of ${\cal S}$ under elongation would lead to a macroscopic surface
charge by Eq.\ (\ref{charge.weyl}).  This is due to the hard-wall boundary
condition, which 
leads to a depletion of negative charge in a layer of thickness 
$\sim \lambda_F$ at the boundary \cite{lang}.  
The macroscopic incompressibility of the electron gas 
can be included by imposing the constraint $\bar{N}_- = \mbox{const.}$
\cite{constraint},  
where $\bar{N}_-$ is given by the first three terms in Eq.\ (\ref{charge.weyl}).
The macroscopic electronic charge $-e\bar{N}_-$ is neutralized by the equal
and opposite positive charge of the jellium background.
The net charge imbalance on the nanocontact (neglecting screening)
is thus $\delta Q_0  =-e\delta N_-$,
which we will show to be quite small---on the order of a single electron charge.
Differentiating Eq.\ (\ref{ener.weyl}) with
respect to $L$ with the constraint $\bar{N}_-=\mbox{const.}$, one finds
\begin{equation}
F=-\left. \frac{\partial \Omega}{\partial L}\right|_{\bar{N}_-}
= -\frac{\sigma_{\cal V}}{5} \frac{\partial {\cal S}}{\partial L}
+ \frac{2}{5} \frac{\partial
({\cal C}/\pi)}{\partial L} \Delta F_{\rm top}
+ \delta F,
\label{force.weyl}
\end{equation}
where $\sigma_{\cal V}=\varepsilon_F k_F^2/16\pi$ is the surface energy of a 
noninteracting electron gas \cite{sbb} at fixed $\cal V$
and $\Delta F_{\rm top}=4\varepsilon_F/9\lambda_F$.
The reduction of the surface energy
by a factor of 5 has been discussed by Lang \cite{lang}.  
The second term on the 
right-hand-side of Eq.\ (\ref{force.weyl}), termed the ``topological force''
by H\"oppler and Zwerger \cite{comment.zwerger} since it depends only on the
topology of the cross-section in the adiabatic limit, is 
reduced by a factor of 2.5.
Importantly, since the 
constraint $\bar{N}_-=\mbox{const.}$ differs from the constraint 
${\cal V}=\mbox{const.}$ used in previous work
\cite{sbb,comment.zwerger,blom,wnw,zwerger2,jerome}
only by terms of order $(k_F D^{\ast})^{-1}$, the 
mesoscopic fluctuations $\delta F$ and $\delta N_-$
are quite insensitive to the choice of constraint.

The fluctuating part of the DOS $\delta g$ 
may be evaluated in the semiclassical (stationary-phase) approximation
as a sum over the periodic classical orbits of the system
\cite{gutzwiller,brack,b&b.wigner,gut.symmetry,gut.interpolate}.
For closed systems, the sum over periodic orbits 
is generically not convergent, and a
broadening of the energy structure in $\delta g(E)$ must be introduced
by hand \cite{brack}.  However, we shall see that for an open system, such
as a nanocontact, the periodic orbit sum converges; the
finite dwell-time of a particle in an open system introduces a natural energy
broadening.

Let us first consider the case of a 2D nanocontact. 
For a finite radius of curvature $R$, 
there is only one unstable periodic classical orbit (plus harmonics),
which moves up and down at the narrowest point of the neck.
One obtains
\begin{equation}
\label{dos.gut.2d}
\delta g^{\rm 2D}_{\rm sc}(E)=
\frac{ 2 mD^* }{\pi \hbar^2 k_E}
\sum_{n=1}^\infty \frac{\cos(2n k_E D^*)}{\sinh(n \chi)},
\end{equation}
where the Lyapunov exponent $\chi$ of the 
primitive periodic orbit satisfies
$\exp(\chi)= 1+D^*/R + \sqrt{(1+D^*/R)^2-1}$.  Eq.\ (\ref{dos.gut.2d})
diverges when $\chi\rightarrow 0$, i.e., when $R\rightarrow \infty$.  In
that limit, the nanocontact acquires translational symmetry along the $z$
axis, so that 
a generalization of the Gutzwiller formula obtained by Creagh and
Littlejohn \cite{gut.symmetry}
must be used, which gives a finite
result.  In this limit, the motion is classically {\em integrable}.
One can treat small deviations from translational symmetry via
perturbation theory in $1/R$.  The resulting
asymptotic behavior for large $R$ may
be combined with the result [Eq.\ (\ref{dos.gut.2d})] valid for small $R$ to 
construct the following interpolation
formula, valid for arbitrary $R$:
\begin{equation} 
\delta g^{\rm 2D}_{\rm int} (E)  = 
\frac{ \sqrt{8} m D^*}{\pi \hbar^2 k_E} 
\sum_{n=1}^\infty \frac{ {\cal C} (2 n k_E D^*- \frac{\pi}{4},
\sqrt{\frac{n k_E L^2}{\pi R}})
}{\sinh(n \chi)},
\label{dos.int.2d}
\end{equation}
where ${\cal C}(x,y)\equiv \cos(x) {\rm C}(y) - \sin(x){\rm S}(y)$, with
C and S Fresnel integrals.  
In Eq.\ (\ref{dos.int.2d}), the specific shape of the nanocontact was taken
to be $D(z)=D^* + z^2/R$.
For a discussion of related interpolation formulae, see Ref.\ 
\onlinecite{gut.interpolate}.
Classically, only the case $R=\infty$ is integrable.  But semiclassically,
there is a smooth crossover between the strongly chaotic ($R\rightarrow 0$) and 
the nearly integrable ($R\rightarrow \infty$) regimes.  The scaling parameter
describing this crossover is
\begin{equation}
\alpha = L/\sqrt{\lambda_F R}.
\label{alpha}
\end{equation}
We refer to $\alpha$ as the {\em quantum chaos parameter}, 
since the quantum fluctuations of
the system correspond to those of a chaotic system when $\alpha \gg 1$ and 
correspond to those of a quasi-integrable system when $\alpha \ll 1$.

Fig.\ \ref{fig.2} shows a comparison of the semiclassical result $g_{\rm sc}=
\bar{g}+\delta g_{\rm int}^{2D}$ and a
numerical calculation of $g$
using a recursive Green's function technique \cite{jerome}.  
The agreement of the semiclassical result and the numerical calculation is
quite good, even in the extreme quantum limit $G \lesssim 2e^2/h$.  The small
discrepancy is of the size expected due to diffractive corrections \cite{brack}
from the sharp corners present in the geometry studied numerically, where
the nanocontact was connected to straight wires of width $k_F D=52$ 
for technical reasons.

The denominator $\sinh n\chi$ in Eqs.\ (\ref{dos.gut.2d}) and 
(\ref{dos.int.2d}) describes 
the effects of tunneling.
In the limit $R\gg D^*$, 
the Lyapunov exponent
$ \chi \rightarrow \sqrt{2D^*/R}$, and one 
recovers the WKB approximation of Ref.\ \onlinecite{sbb}. 
In the 
opposite limit $R\ll D^*$, 
$\sinh \chi \rightarrow D^*/R$, so
$\delta g$ is suppressed relative to the
value expected in the WKB approximation (which neglects
tunneling) by a factor of  
$\sqrt{2R/D^*}$.
In the adia-
\noindent
batic approximation, the energies of the transverse modes
in the point contact are
$\varepsilon_n(z) = (\hbar^2/2m) (\pi n/D(z))^2
= \varepsilon_n(0) -  m\omega_n^2 z^2/2 + \cdots$
and the probability that an electron of energy $E$ in mode $n$ will be
transmitted through the point contact is \cite{beenakker}
$T_n(E) \simeq \left(1+\exp\left\{-2\pi[E-\varepsilon_n(0)]/\hbar
\omega_n\right\}\right)^{-1}$.
The quality of the conductance quantization thus decreases {\em exponentially}
with the parameter 
$\hbar \omega_n/\Delta \varepsilon_n \simeq 
\pi^{-1} \sqrt{2D^*/R}$,
where $\Delta \varepsilon_n = \varepsilon_n - \varepsilon_{n-1}$,
while 
the DOS fluctuations $\delta g$ are suppressed only {\em inversely proportional}
to this parameter.
The fact that the suppression in
each case depends only on the ratio $D^*/R$ implies that the suppression
of $\delta g$, like the degradation of conductance quantization, is
a consequence of {\em tunneling}.  Indeed, it is the rounding of the DOS
due to tunneling that causes the sums over $n$ in Eqs.\ 
(\ref{dos.gut.2d}) and (\ref{dos.int.2d}) to {\em converge}.
That quantum tunneling through a point 
contact can be expressed 
purely in terms
of the instability $\chi$ of the classical periodic orbits within the contact
is remarkable.  

Let us now consider the experimentally relevant case of an 
axially-symmetric 3D nanocontact.  For finite $R$, all classical periodic
orbits lie in the plane of the narrowest cross section of the contact; however
there are now countably many distinct families of singly-degenerate 
periodic orbits \cite{brack,zwerger2},
labeled by their winding number $w$ about the axis of symmetry $z$ and by
the number of vertices $v\geq 2w$.
The interpolation formula for $\delta g$, describing the crossover from the 
chaotic regime $\alpha \gg 1$ 
to the integrable regime $\alpha \ll 1$, 
is
\begin{eqnarray}
\delta g^{\rm 3D}_{\rm int}(E) & = &
\frac{ m}{\hbar^2 \sqrt{\pi k_E}} \sum_{w=1}^\infty \sum_{v=2w}^\infty 
\frac{f_{vw} L_{vw}^{3/2}}{v^2 \sinh(v \chi_{vw}/2)}
\nonumber \\
& & \times
{\cal C}(k_E L_{vw} - {\textstyle \frac{3 v \pi}{2}}, 
\alpha \sqrt{v\sin\phi_{vw}k_E/k_F}),
\label{dos.int.3d}
\end{eqnarray}
where 
$\phi_{vw}=\pi w/v$, 
$f_{vw} = 1+\theta(v-2w)$,
and
\[
\exp(\chi_{vw}) = 
1+\frac{L_{vw} \sin \phi_{vw}
}{vR}+\sqrt{\left(1+\frac{ L_{vw} \sin \phi_{vw}}{vR}\right)^2-1},
\]
with $L_{vw}=v D^* \sin \phi_{vw}$ the length of a periodic orbit.
We emphasize that the double sum over $w$ and $v$ in
Eq.\ (\ref{dos.int.3d}) converges due to the finite Lyapunov exponent
$\chi_{vw}$.  In Eq.\ (\ref{dos.int.3d}), higher-order terms in the 
small parameter $1/k_F D^*$ ($\mbox{}<0.21$ for contacts of nonzero
conductance) have been omitted.

The mesoscopic force and charge fluctuations are calculated by inserting
Eq.\ (\ref{dos.int.3d}) into Eqs.\ (\ref{ener.total}), (\ref{charge.total})
and (\ref{force.weyl}).  
In order to demonstrate the {\em universality} of the force oscillations, 
it is necessary to make some physically reasonable assumptions regarding the
scaling of the geometry when the nanowire is elongated.  It is natural to 
assume that the deformation occurs predominantly in the narrowest section, 
where the wire is weakest.  This assumption, combined with the constraint of
incompressibility $\bar{N}_-=\mbox{const.}$, implies
$D^{*2} L \approx \mbox{const.}$ 
Furthermore, the radius of curvature
$R\propto L^2/(D-D^*)$,  where $D$ is the diameter at $\pm L/2$, which implies
$\partial \ln R/ \partial \ln L=2+
(\partial \ln D^*/ \partial \ln L)/(D/D^*-1)
\approx 2$.
Thus the quantum chaos parameter 
$\alpha 
\approx {\rm const.}$ under elongation.

Using these assumptions about the scaling of the geometry with elongation,
the derivative with respect to $L$ in Eq.\ (\ref{force.weyl}) can be
evaluated; 
the general formula for $\delta F$ 
is rather lengthy, and will be presented elsewhere.
Here we give only the limiting behavior of the leading-order semiclassical
results:
\begin{eqnarray}
\delta F 
&\begin{array}{c}\mbox{ }\\ \simeq \\ {\scriptstyle \alpha \gg 1} \end{array}&
\frac{\varepsilon_F}{L} \sum_{w=1}^{\infty}
\sum_{v=2w}^{\infty} \sqrt{\frac{L_{vw}}{\lambda_F}} \frac{f_{vw}
\sin(k_F L_{vw} - b_v)}{v^2\sinh(v\chi_{vw}/2)}, 
\label{trace.f1}\\
\delta F &  \begin{array}{c} \mbox{ }\\ \simeq  \\ {\scriptstyle \alpha \ll 1}
 \end{array}& 
-\frac{2\varepsilon_F}{\lambda_F} \sum_{w=1}^{\infty} \sum_{v=2w}^{\infty}
\frac{f_{vw}}{v^2} \sin(k_F L_{vw} - 3v\pi/2), 
\label{trace.f2}
\end{eqnarray}
where $b_v=3v\pi/2 - \pi/4$.   
$\delta F$ is an oscillatory
function of $k_F D^*$; the conductance of the contact is also determined by
$k_F D^*$, indicating that the force oscillations are synchronized with the
conductance steps, as shown in Ref.\ \onlinecite{sbb} and observed 
experimentally \cite{nanoforce}.

The rms amplitude of the force oscillations may be readily calculated 
from Eqs.\ (\ref{trace.f1}) and (\ref{trace.f2}).
We find that $\mbox{rms}\,\delta F$ is independent of $D^*$,
and, apart from small corrections due to tunneling when $R\ll D^*$,
depends only on the quantum chaos parameter $\alpha$:
\begin{equation}
\mbox{rms}\,\delta F = \left\{\begin{array}[c]{c c} 
0.36208 \, \alpha^{-1} \, {\displaystyle \frac{\varepsilon_F}{\lambda_F}}, & 
\alpha \gg 1, \\
&\\
0.58621 \, {\displaystyle \frac{\varepsilon_F}{\lambda_F}}, & 
\alpha \ll 1.
\end{array}\right.
\label{universality}
\end{equation}
The result for $\alpha \ll 1$ agrees with the result for a
straight wire ($\alpha=0$) derived previously by
H\"oppler and Zwerger \cite{zwerger2}.
Eq.\ (\ref{universality}) is also consistent with
previous results based on the WKB approximation \cite{sbb}.
For a realistic geometry of the nanowire \cite{nanoforce},
one expects both the radius of curvature and the elongation to be on the 
scale of $\lambda_F$, implying $\alpha \sim 1$.
There is also experimental evidence \cite{monatomic} of
exceptional geometries with $R\gg \lambda_F$, implying $\alpha \ll 1$.
Thus the mesoscopic oscillations
of the cohesive force are expected to be universal
$\mbox{rms}\,\delta F \sim \varepsilon_F/\lambda_F\approx 1nN$ in monovalent
metals, in agreement with all available experimental data
\cite{nanoforce}.

In nanowires lacking axial symmetry, 
e.g., with an
aspect ratio $a \gg 1$, one can show that $\mbox{rms}\,\delta F \sim
a \varepsilon_F/\lambda_F$.  However, such shapes are energetically 
highly unfavorable due to the increased surface energy.  Eq.\ 
(\ref{universality}) is therefore expected to describe all 
spontaneously occuring nanocontacts.

Eq.\ (\ref{dos.int.3d}) and the assumption 
$D^{*2} L = \mbox{const.}$ 
imply that the force and
charge oscillations are proportional to each other in 3D nanocontacts:
$\delta F = -\varepsilon_F \delta N_-/L
+{\cal O}(1/k_F D^*)$.
In an interacting system, the charge oscillations 
are screened \cite{wnw}, and
the Hartree correction to the grand canonical potential 
is bounded by $\Delta \Omega 
< \delta N_-^2/2g(\varepsilon_F)$.
Evaluating the elementary sums over periodic orbits, we find that the 
average interaction correction $\langle \Delta \Omega\rangle$ 
is small compared to the 
mesoscopic oscillations of $\Omega$:
\begin{equation}
\frac{\langle \Delta \Omega \rangle}{\mbox{rms}\,\delta \Omega} <
\frac{1.36791}{k_F D^*},
\label{interact}
\end{equation}
where $k_F D^* > 4.81$ for a contact with nonzero conductance.  
This result justifies the use of the independent-electron approximation 
\cite{sbb,comment.zwerger,blom,wnw,zwerger2,jerome}.

In conclusion, we have shown that trace formulae \`a la Gutzwiller 
converge and give quantitatively accurate results for 
the equilibrium quantum fluctuations in point contacts and nanowires.
Using this approach, we have shown that the cohesive
force of a metallic nanocontact, modeled as a hard-wall
constriction in an electron gas, exhibits universal mesoscopic oscillations
whose size $\mbox{rms}\, \delta F \sim \varepsilon_F/\lambda_F$ is 
independent of the conductance and shape of the contact, and depends only
on a dimensionless parameter $\alpha$ characterizing the degree of 
quantum chaos.  Our prediction of universality is consistent with all
experiments performed to date \cite{nanoforce}.

%                          Acknowledgements

We wish to thank R.\ Bl\"umel, J.\ Morris, R.\ Prange, and H. Primack
for useful discussions.  This
work was supported in part by the National Science Foundation under 
Grant No.\ PHY94--07194.  F.\ K.\ acknowledges support from grant SFB 276
of the Deutsche Forschungsgemeinschaft.
J.\ B.\ acknowledges support from Swiss National Foundation PNR 36 
``Nanosciences'' grant \# 4036-044033.


\begin{references}

\bibitem{kac} M. Kac, Am. Math. Monthly {\bf 73}, 1 (1966).

\bibitem{yanez}
G. Guti\'erez and J. M. Y\'a\~nez, Am. J. Phys. {\bf 65}, 739 
(1997).

\bibitem{nanoforce} C. Rubio, N. Agra\"{\i}t, and S. Vieira, Phys. Rev. Lett.
{\bf 76}, 2302 (1996); 
A. Stalder and U. D\"urig, Appl. Phys. Lett. {\bf 68},
637 (1996); 
C. Untiedt, G. Rubio, S. Vieira, and N. Agra\"{\i}t,
Phys. Rev. B {\bf 56}, 2154 (1997).

\bibitem{sbb} C. A. Stafford, D. Baeriswyl, and J. B\"urki, Phys. Rev. Lett.
{\bf 79}, 2863 (1997).


\bibitem{comment.zwerger} C. H\"oppler and W. Zwerger, Phys. Rev. Lett.
{\bf 80}, 1792 (1998).

\bibitem{blom} S. Blom et al., 
Phys. Rev. B {\bf 57}, 8830 (1998).

\bibitem{landman2}
C. Yannouleas, E. N. Bogachek, and U. Landman, Phys. 
Rev. B {\bf 57}, 4872 (1998).


\bibitem{wnw} F. Kassubek, C. A. Stafford, and H. Grabert, Phys. Rev.  B
{\bf 59}, 7560 (1999).

\bibitem{zwerger2} C. H\"oppler and W. Zwerger, Phys. Rev. B {\bf 59},
R7849 (1999).

\bibitem{jerome} J. B\"urki, C. A. Stafford, X. Zotos, and 
D. Baeriswyl, cond-mat/9903006 (preprint). 


\bibitem{beenakker} For a review, see
C. W. J. Beenakker and H. van Houten, {\em Solid State Physics}
{\bf 44}, 1 (1991).


\bibitem{gutzwiller} M. C. Gutzwiller, {\em Chaos in Classical and 
Quantum Mechanics} (Springer Verlag, New York, 1990).

\bibitem{brack} 
M. Brack and R. K. Bhaduri,
{\em Semiclassical Physics}, 
Frontiers in Physics, Vol. 96 (Addison Wesley, 1997).


\bibitem{b&b.wigner} R. Balian and C. Bloch, Ann. Phys. (N.Y.)
{\bf 85}, 514 (1974).

\bibitem{gut.symmetry} S. C. Creagh and R. G. Littlejohn, Phys. Rev. A
{\bf 44}, 836 (1991).


\bibitem{gut.interpolate} D. Ullmo, M. Grinberg, and S. Tomsovic,
Phys. Rev. E {\bf 54}, 136 (1996).




\bibitem{domega}
In Eq.\ (\ref{ener.total}), we have neglected
a contribution $\Delta \Omega$ due to 
electron-electron interactions,  
which will be shown to be
unimportant for mesoscopic effects.


\bibitem{dashen} R. Dashen, S.-K. Ma, and H. J. Bernstein, Phys. Rev.
{\bf 187}, 345 (1969).




\bibitem{lang} N. D. Lang, {\em Solid State Physics}
{\bf 28}, 225 (1973).

\bibitem{constraint}
One should not impose the constraint
$N_- =\mbox{const.}$, 
which would require the positive background to be infinitely
soft, to adapt to every mesoscopic variation in the electron
density.


\bibitem{monatomic} H. Ohnishi, Y. Kondo, and K. Takayanagi, Nature
{\bf 395}, 780 (1998); A. I. Yanson et al., 
ibid. {\bf 395}, 783 (1998).


\end{references}
\end{document}